\documentclass[aip,reprint]{revtex4-1}
\usepackage[pdftex]{graphicx} 
\usepackage{float}
\usepackage{amsmath,color}
\usepackage{subcaption}

\draft 

\begin{document}


\title{Guide Field Effects on the Distribution of Plasmoids in Multiple Scale Reconnection} 



\author{Stephen Majeski}
\email[]{smajeski@princeton.edu}
\author{Hantao Ji}
\author{Jonathan Jara-Almonte}
\author{Jongsoo Yoo}
\affiliation{Princeton Plasma Physics Laboratory, Princeton University, Princeton, New Jersey 08544 USA}


\date{\today}

\begin{abstract}
The effects of a finite guide field on the distribution of plasmoids in high-Lundquist-number current sheets undergoing magnetic reconnection in large plasmas are investigated with statistical models. Merging of plasmoids is taken into account either assuming that guide field flux is conserved resulting in non-force-free profiles in general, or that magnetic helicity is conserved and Taylor relaxation occurs to convert part of the summed guide field flux into reconnecting field flux towards minimum energy states resulting in force-free profiles. It is found that the plasmoid distribution in terms of reconnecting field flux follows a power law with index 7/4 or 1 depending on whether merger frequencies are independent of or dependent on their relative velocity to the outflow speed, respectively. This result is approximately the same for the force-free and non-force-free models, with non-force-free models exhibiting indices of 2 and 1 for the same velocity dependencies. Distributions in terms of guide field flux yield indices of 3/2 for the non-force-free model regardless of velocity dependence. This is notably distinct from the indices of 11/8 and 1 for the force-free models independent of and dependent on velocity, respectively. At low guide field fluxes the force-free models exhibit a second power law index of 1/2 due to non-constant flux growth rates. The velocity dependent force-free model predicts the production of slightly more rapidly moving large guide field flux plasmoids which is supported by observational evidence of flux ropes with strong core fields. Implications are discussed on particle acceleration via Fermi processes.
\end{abstract}


\maketitle 


\section{Introduction}

Impulsively fast magnetic reconnection is a ubiquitous phenomenon widely observed in magnetized space, solar, astrophysical, and laboratory fusion plasmas~\cite{yamada10,ji11,lin_cranmer_farrugia_2008,zweibel_yamada_2009}. However, fast collisionless reconnection mechanisms based on non-MHD (Magnetohydrodynamic) effects such as two-fluid or kinetic effects~\cite[e.g.][]{birn01} are only applicable to scales much smaller than system sizes of these magnetized plasmas in space and astrophysics. In large systems, collisional MHD models such as the Sweet-Parker model are commonly used to describe reconnection processes. However, the predicted Sweet-Parker reconnection rate is much too slow to be consistent with many large scale phenomena like solar flares. At high Lundquist numbers, it has been realized that these current sheets can potentially fracture in a cascading process which results in a significantly increased reconnection rate~\cite{shibata_tanuma_2001}. The instability responsible for this process is the plasmoid instability, which can grow rapidly leading to a reconnection rate that remains fast and is weakly dependent on the Lundquist number~\cite{loureiro_schekochihin_cowley_2007,bhattacharjee_huang_yang_rogers_2009,comisso_2017,comisso_lingam_2016}. Thus, the plasmoid instability has been proposed as a promising mechanism to couple global system scale to local dissipation scales in reconnecting current sheets, although there exist alternative models based on MHD turbulence\cite[e.g.][]{lazarian99}.

In the MHD regime this instability is found at sufficiently high Lundquist numbers~\cite{loureiro_schekochihin_cowley_2007}, causing the breakup of long current sheets into chains of self contained magnetic islands. These magnetic islands are highly dynamic, interacting with each other as they move through the length of the primary current sheet until they exit. Electrons can be efficiently accelerated to high energies by these dynamic magnetic islands~\cite{drake_swisdak_che_shay_2006,sironi2014} emitting observable radiation~\cite[e.g.][]{nalewajko_uzdensky_2015}. The acceleration process is based on reflection of particles from the ends of each contracting magnetic island resulting in first order Fermi acceleration. Additionally, higher energy particles which are free to move throughout the primary current sheet across multiple islands can be mirrored and accelerated by an enhanced Fermi process~\cite{hoshino12,hoshino_lyubarsky_2012}. A better understanding of the magnetic configuration of plasmoids and their statistical properties may help reveal how plasmoids can contract, the relative velocities with which they interact, and hence how the particles within and in between are accelerated.

Previous studies have shown how this instability can interrupt a Sweet-Parker current sheet which surpasses $S_c \sim 10^4$ depending on preexisting noise~\cite{huang10,comisso_2017}. A plasmoid unstable current sheet undergoes a cascading process which generates new plasmoids until their mean separation is low enough that the local Lundquist number of reconnecting current sheets between them has been reduced below $S_c$, thus stable to the plasmoid instability~\cite{shibata_tanuma_2001}. These plasmoids, however, are highly dynamic through their interaction. Between the time when a plasmoid is born and the time when it advects out of the reconnection layer, it may absorb smaller plasmoids or be absorbed by larger plasmoids. Reconnection of their surrounding current sheets leads to the accretion of flux as well. Over time, the current sheet may produce very large plasmoids on the order of one tenth the length of the sheet or greater\cite{lssu_2012}.

Due to the importance of plasmoids in reconnection adjacent phenomena, statistical scalings have been sought both analytically and numerically to describe the distribution of plasmoids most commonly in terms of their reconnecting flux $\psi$, as $f(\psi)$. In the Hall-MHD (magnetohydrodynamic) regime, $f \propto \exp(-\psi)$ behavior has been predicted~\cite{fermo10}. This exponential dependence of reconnecting flux has been observed in the near-Earth space with limited \textit{in-situ} measurements by a few satellites~\cite{fermo11,bergstedt20}, by remote-sensing measurements of solar eruptions~\cite{guo13}, and in laboratory experiments~\cite{dorfman14,olson16} but with limited resolutions. In the 3D MHD regime without a guide field, an entropy variational principle has been used to derive~\cite{lingam_comisso_2018} a power law index of 3, while 2D relativistic particle-in-cell simulations have been paired with Monte Carlo methods to uncover a power law index in the range of 1 to 2. The often reported scaling is $f(\psi) \propto \psi^{-2}$, seen in many 2D MHD simulations and justified in most cases by statistical approaches~\cite{uzdensky10,Loureiro2012,huang_bhattacharjee_2012,takamoto13,shen13}. It has also been argued that the combination of a power law index of 1 with an exponential tail could appear as a power law index of 2 in numerical results, and analytical models explaining this behavior have been proposed~\cite{huang_bhattacharjee_2012,guo13}. Specifically, Huang et al~\cite{huang_bhattacharjee_2012} developed a statistical approach which deftly demonstrates how the power law dependence of the distribution can be replicated by accounting for the essential behaviors of the unstable current sheet. Two models, one which allows for variation in the relative velocity to the mean outflow of these plasmoids and one which does not, produce power law indices of 1 and 2 of the reconnecting flux, respectively. All of these models do not explicitly take into account of the presence of a finite guide field in the reconnecting current direction. In many natural circumstances, however, a finite guide field is present during reconnection which may modify the distribution of plasmoids. Work by Ni et al \cite{ni_lin_murphy_2013} found that the presence of a guide field does not have a major impact on the instability itself. We expand here the approach of Huang et al. to investigate effects of a finite guide field on the plasmoid distributions.

Both models of Huang et al.~\cite{huang_bhattacharjee_2012} are further developed here by adopting force-free field profiles internal to plasmoids as a result of Taylor relaxation~\cite{taylor86} to determine the distribution of plasmoids in the presence of a finite guide field. Without a guide field, the current sheet and the plasmoids within are essentially non-force-free, \textit{e.g.} the incoming reconnecting field pressure is balanced by plasma pressure. With a finite guide field, however, plasmoids become magnetic flux ropes~\cite{moldwin91} which are long known~\cite{slavin95} to relax towards a force-free state possessing a strong core field in an approximately cylindrical shape~\cite{Slavin03jgr}. Such force-free fields are also known in the laboratory pinch experiments as a result of Taylor relaxation~\cite{taylor74,taylor86}, during which magnetic energy is minimized while conserving magnetic helicity~\cite{ji95a}. 
The force-free fields, $\vec{B}$, are given by~\cite{taylor74,chandrasekhar_kendall_1957}
\begin{equation}\label{eq:eval}
    \nabla \times \vec{B} = \lambda \vec{B},
\end{equation}
where $\lambda$ is the eigenvalue determined by boundary conditions. The lowest order solution to this equation was found to be $B_z = B_0 J_0(\lambda r)$ and $B_{\theta} = B_0 J_1(\lambda r)$ in cylindrical coordinates. Here $J_0$ and $J_1$ are the zeroth and first order Bessel functions of the first kind, respectively. Taylor showed~\cite{taylor74} that in toroidal pinch experiments with large Lundquist numbers and moderate toroidal fields, these profiles reasonably predicted the peaked toroidal (core) field at the center with a much reduced magnitude or even reversed direction at the edge by extending beyond the first zero of $J_0$.

Magnetic reconnection with a guide field can be thought of as transporting magnetic helicity from the background into the plasmoids in the current sheet. If these plasmoids with a finite magnetic helicity are allowed to relax towards a minimum energy state while advecting towards the current sheet exit, their internal field structures will take a form of force-free profiles. When plasmoids merge, magnetic helicity can be dissipated in the secondary current sheets. However for high Lundquist numbers ($\geq S_c$) this dissipation is low for moderate to weak guide fields~\cite{ji_1999} while magnetic energy is reduced significantly during merging. Therefore, the plasmoid merging process can be also regarded as a Taylor relaxation process to minimize magnetic energy while conserving the total magnetic helicity so that the resultant plasmoids also take the form of force-free profiles. Making use of these assumptions, the anti-parallel reconnection model is modified to include the effects of Taylor relaxation with a finite but moderate guide field, following plasmoid mergers where adequate time for relaxation is available. These details are discussed in Sections \ref{sec:dist} and \ref{sec:disc}. Hence, a distribution of reconnecting and guide field plasmoid fluxes can be obtained. An alternative model is also provided which simply adds guide field fluxes upon plasmoid coalescence without Taylor relaxation~\cite{zhou2020}. This represents the strong guide field regime where Taylor relaxation is not allowed and the plasmoids are not force-free while non-negligible magnetic helicity is dissipated in the merging secondary current sheets. For both the relative-velocity independent and velocity dependent models, the non-force-free and force-free distributions are compared and their implications and differences are discussed. 

In Section \ref{sec:flxsln} we derive the relationships necessary to relate the reconnecting and guide field fluxes of a plasmoid to its magnetic helicity, providing the rate at which plasmoids gain magnetic helicity from the background reconnection. Some assumptions of the cylindrical plasmoid model are also discussed, along with their validity regimes. Section \ref{sec:dist} lays out further assumptions about magnetic helicity dissipation as relevant to the construction of a statistical equation which conserves magnetic helicity in plasmoid mergers. The non-force-free, guide field conserving model is also provided as the alternate regime of a strong guide field. Section \ref{sec:nums} provides the numerical methods and solutions to the equations derived in Section \ref{sec:dist}. These distributions and their features are discussed in detail in Section \ref{sec:disc}. With solutions in hand, the assumptions made earlier can be verified or constrained. Section \ref{sec:conc} summarizes the results, especially where relevant to particle acceleration. Possible future improvements to the statistical equations are also discussed.

\section{Theoretical Models}

\subsection{Relations between Relaxed Plasmoids and the Reconnecting Current Sheet}\label{sec:flxsln}

\begin{figure}
\centering{
  \begin{subfigure}{\linewidth}
    \includegraphics[width=.8\linewidth]{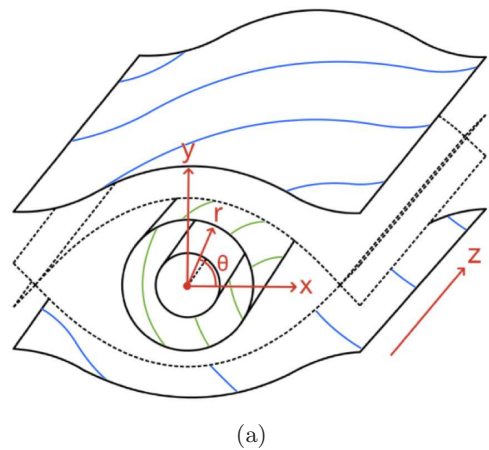}
    \caption{ }
  \end{subfigure}
  \begin{subfigure}{\linewidth}
    \includegraphics[width=.85\linewidth]{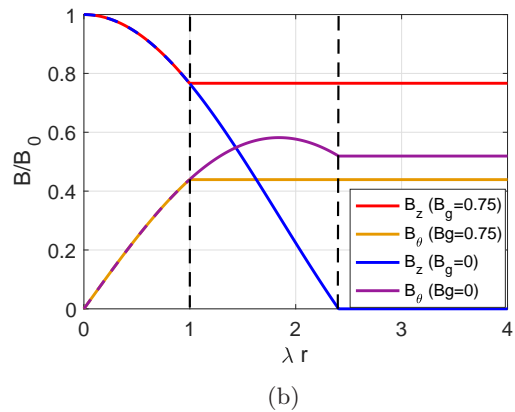}
    \caption{ }
  \end{subfigure}
  \begin{subfigure}{\linewidth}
    \includegraphics[width=.85\linewidth]{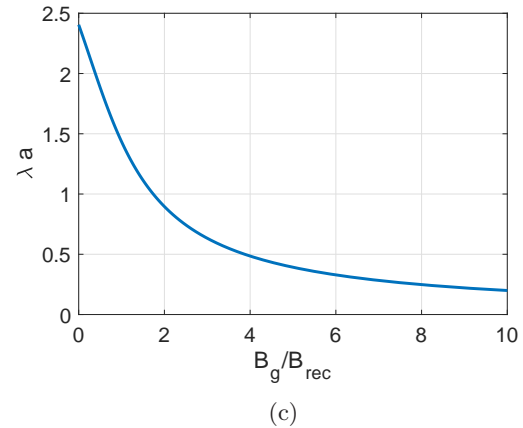}
    \caption{ }
  \end{subfigure}
 }
\caption{(a) Schematic of a cylindrical plasmoid embedded in part of the reconnecting current sheet shown with the coordinates used in this paper. The dotted separatrix coincides with either $\lambda a \approx 2.405$ or $\approx 1$ in (b). Field lines are drawn on the select surfaces (omitting the separatrix and innermost surface), shown as blue outside the plasmoid and green inside. We assume for simplicity the field lines are nearly circular up until the separatrix. (b) The plasmoid field profiles are shown to connect to the constant background field at $r=a$, for the cases of $B_g/B_{rec}=0.77$ and $B_g/B_{rec}=0$. (c) The functional dependence of $\lambda a$ versus $B_g/B_{rec}$ is shown to illustrate their relationship.}
\label{fig:taylorprofs}
\end{figure}

Several simple but important relations between force-free plasmoids and the reconnecting current sheet where they are embedded are described in this Section. Each isolated plasmoid either when they are born or as a result of merging of two plasmoids is assumed to take the force-free profiles, as predicted by Taylor through minimizing magnetic energy while conserving magnetic helicity. Unlike Taylor, however, both the guide field and reconnecting field fluxes are not conserved; rather they are determined by the embedded current sheet as follows. We assume that these plasmoids are simple axisymmetric cylinders as shown in Fig.~\ref{fig:taylorprofs}(a) embedded in the reconnecting current sheet. The relations between these plasmoids and the background current sheet are derived in a straightforward manner by relating the plasmoid radius $a$ to the eigenvalue $\lambda$. At the edge of the plasmoid $r=a$, we match the background reconnecting and guide fields (termed $B_{rec}$ and $B_g$, respectively) to the azimuthal and axial magnetic components of the cylindrical plasmoids, $B_{\theta}$ and $B_z$ respectively.  We note that this is less valid for smaller, newly born plasmoids as their radius is not sufficiently large to reach the asymptotic values of $B_{rec}$ and $B_g$. The impact of this assumption, however, is likely small, and will be discussed in Section \ref{sec:nums}. Eliminating $B_0$ from the profiles, we have that
\begin{equation}\label{eq:bgnd}
    \frac{B_{rec}}{B_g} = \frac{J_1(\lambda a)}{J_0(\lambda a)}.
\end{equation}

There are infinitely many solutions to this equation. However, we assume that the solution with the lowest value of $\lambda a$ should be used. This is because solutions with higher $\lambda a$ involve a reversing guide field, yielding higher energy configurations. These configurations are unstable to kink modes past the critical $\lambda a = 3.176$, leading to non-axisymmetric dynamics which we do not seek to capture in this model.~\cite{voslamber1962} This assumption is supported by observations which do not typically exhibit a reversed guide or core field within plasmoids. For the rest of this paper, the lowest solution with $\lambda a$ satisfying Eq.~(\ref{eq:bgnd}) will be used. Figure~\ref{fig:taylorprofs}(b) shows two such examples of lowest $\lambda a$: one for $B_g=0.75B_{rec}$ and one for $B_g=0$. For each choice of $B_g/B_{rec}$, the corresponding $\lambda a$ is determined as shown in Fig.~\ref{fig:taylorprofs}(c). Furthermore, once $\lambda a$ is determined, $B_0$ may also be determined from both $B_g = B_0 J_0(\lambda a)$ and $B_{rec} = B_0 J_1(\lambda a)$.

The reconnecting flux per unit length ($\psi$) and the guide field flux ($\phi$) contained within a plasmoid will be defined as follows:
\begin{equation}\label{eqn:fluxes}
    \psi = \int_0^a B_0 J_1(\lambda r) dr, \quad \phi = \int_0^a B_0 J_0(\lambda r) 2\pi rdr.
\end{equation}
For a toroidal system, $\psi$ is the analog of the poloidal flux per unit axial length and $\phi$ is the analog of the toroidal flux. In the construction of our statistical kinetic equation it will become necessary to relate the reconnecting flux, guide field flux, and magnetic helicity through the background quantities by eliminating individual dependencies on $\lambda$ and $a$. The eigenvalue equation can be integrated over an area with the $\hat{z}$ normal, which using Stoke's theorem and Eq.~(\ref{eqn:fluxes}) gives:
\begin{equation}\label{eqn:z}
\begin{gathered}
    \int \nabla \times B \cdot \hat{z}dA = \oint B_{\theta}(a)\hat{\theta} \cdot d\vec{\textit{l}} = \\
    2\pi a B_{rec} = \lambda \int \vec{B} \cdot \hat{z}dA = \lambda\phi.
\end{gathered}
\end{equation}
Performing the same integration with the $\hat{\theta}$ normal we obtain:
\begin{equation}\label{eqn:theta}
\begin{gathered}
    \int \nabla \times B \cdot \hat{\theta}dA = \oint B_z\hat{z} \cdot d\vec{\textit{l}} = \\ 
    (B_0-B_g)l = \lambda \int \vec{B} \cdot \hat{\theta}dA = \lambda \psi l.
\end{gathered}
\end{equation}
The loop integration was performed along the magnetic axis for a distance of $l$ and returns on the cylindrical surface where the axial magnetic field component equals the background $B_g$. There is no contribution to the loop integration at the ends of the plasmoid due to the radial magnetic field being zero in the model.
Dividing Eq.~(\ref{eqn:z}) by the square of Eq.~(\ref{eqn:theta}) with rearrangements leads to
\begin{equation}\label{eqn:psiphi}
    \frac{2\pi \lambda a B_{rec} B_0}{(B_0-B_g)^2} = \frac{B_0 \phi}{\psi^2} = \zeta,
\end{equation}
where the constant dimensionless term has been renamed as $\zeta$ for simplicity. The same conclusion can be arrived at from dimensional analysis when the goal is an expression relating $\psi$ and $\phi$ which does not expressly contain statistical variables such as the plasmoid radius $a$. Instead, only the combination of $\lambda a$ is involved as specified in Fig.1(c) for a given $B_g/B_{rec}$. Using $\nabla \times \vec{A} = \vec{B}$ the vector potential can be found for each plasmoid, namely:
\begin{equation}
\vec{A}(\lambda r) = \frac{B_0}{\lambda}\left[\left[J_1(\lambda r)-\frac{a}{r}J_1(\lambda a)\right]\hat{\theta} + J_0(\lambda r)\hat{z}\right],
\end{equation} 
where the gauge choice has been made such that $A_{\theta}(\lambda a)=0$ for each plasmoid to be isolated to avoid generating magnetic helicity through linking neighboring plasmoids and surroundings. (Otherwise the concept of relative helicity~\cite{berger1984,finn1985} needs to be invoked.) The magnetic helicity for each plasmoid may then be calculated directly using $K = \int \vec{A}\cdot\vec{B}d^3r$. The resulting relation can be manipulated to yield
\begin{equation}
    K = \psi\phi \biggl[\lambda a\frac{B_{rec}^2+B_g^2}{B_{rec}(B_0-B_g)} + \frac{B_0}{B_g-B_0}\biggr] = \alpha \psi \phi,
\end{equation}
where the constant coefficient has been renamed $\alpha$. This gives a simple relationship for the magnetic helicity which depends only on $\psi$ and the background parameters:
\begin{equation}
    K = \alpha(\zeta/B_0) \psi^3.
\label{helicity}
\end{equation}

The growth rate of the plasmoid magnetic helicity due to reconnection if the plasmoid relaxes more rapidly than the rate by which the primary current sheet reconnects (an assumption to be justified in Section \ref{sec:dist}), is given by
\begin{equation}\label{eq:dkdt}
    \frac{dK}{dt} = 3\gamma (\alpha(\zeta/B_0))^{1/3}K^{2/3}.
\end{equation}
The assignment $d\psi/dt = \gamma$ has been made for brevity. To determine the final state of a plasmoid which has formed as the result of the merger of two plasmoids, we may then exploit either guide field flux conservation or magnetic helicity conservation depending on the background conditions. 


\subsection{The Kinetic Equations for the Plasmoid Distribution}\label{sec:dist}

A standard governing kinetic equation for plasmoid size distribution in terms of the reconnecting field flux includes contributions from the sources, growth, and sinks of plasmoids as they evolve in the dynamic current sheets. The source of plasmoids is always localized at zero flux while their growth is based on the overall reconnection rate. The latter is determined by the critical Lundquist number for a Sweet-Parker current sheet to become unstable~\cite{bhattacharjee_huang_yang_rogers_2009}: $S_c \sim 10^4$. Therefore, the stable current sheets which lie in between plasmoids add flux to the plasmoids at a rate $d\psi/dt=BV_A/\sqrt{S_c}=\gamma$. Following Huang \& Bhattacharjee~\cite{huang_bhattacharjee_2012}, with the assumption that mergers occur with a frequency independent of the plasmoid relative velocities, the statistical kinetic equation is given by
\begin{equation}\label{eq:hb}
    \frac{\partial f}{\partial t} + \gamma \frac{\partial f}{\partial \psi} = \xi \delta (\psi) - \frac{n_>}{\tau_A}f - \frac{1}{\tau_A}f,
\end{equation}
where
\begin{equation}
    n_>(\psi) = \int_{\psi}^{\infty} f(\psi') d\psi'.
\end{equation}
All of the solutions obtained are steady state with $\partial_t = 0$. On the right, the first term, $\xi \delta (\psi)$, represents the creation of plasmoids where $\delta (\psi)$ is Dirac $\delta$-function. The second term represents the loss due to absorption by larger plasmoids with a rate proportional to their number, $n_>(\psi)$, assuming a typical relative speed on the order of $V_A$. The third is the sink term representing the advection of plasmoids out of the reconnection layer. The Alfv\'en time $\tau_A = L/V_A$ is that of the reconnecting component of the background current sheet. The solution for this equation was determined to be~\cite{huang_bhattacharjee_2012}
\begin{equation}
    f(\psi) = \frac{2C}{\gamma\tau_A} \frac{\exp(-\psi/\gamma \tau_A)}{[C-\exp(-\psi/\gamma \tau_A)]^2}
\end{equation}
where $C = 1+2/N$ and $N$ is the number of plasmoids. This solution possesses three regimes: a constant initial section where plasmoids grow due to reconnection not having had time to merge yet, followed by a power law dependence with the index of 2 where mergers dominate dynamics, then lastly an exponential tail where advection loss dominates for the largest plasmoids. When accounting for plasmoids with varying velocities relative to the mean outflow, the statistical kinetic equation takes on a modified merger rate according to:
\begin{equation}\label{eq:hbv}
    \frac{\partial F}{\partial t} + \gamma \frac{\partial F}{\partial \psi} = \xi \delta (\psi) h(v) - \frac{H}{\tau_A}F - \frac{1}{\tau_A}F,
\end{equation}
where
\begin{equation}
    H(\psi,v) = \int_{\psi}^{\infty} \int_{-\infty}^{\infty} \frac{|v-v'|}{V_A} F(\psi',v') dv' d\psi'.
\end{equation}

The plasmoid creation term is now endowed with a velocity distribution, assumed to be $h(v) = (1/\sqrt{\pi}V_A)\exp(-v^2/V_A^2)$ (although as noted by Huang et al~\cite{huang_bhattacharjee_2012} and confirmed by our own tests the solution is not very sensitive to the exact distribution chosen). The coalescence term corrects the rate at which mergers occur by the difference in velocity between the two merging plasmoids, approximating the resultant velocity as that of the larger plasmoid. An analytical solution to the steady state form of this equation is not known, it is instead solved numerically.

\subsection{A Model for Force-Free Plasmoids}

Equations \ref{eq:hb} and \ref{eq:hbv} can be modified to account for the effects of Taylor relaxation with a guide field. As the reconnecting flux of a plasmoid grows by reconnection, its magnetic helicity also grows. This change in magnetic helicity calls for a continuous modification to the plasmoid internal profiles if there is sufficient time for the plasmoid to continuously relax. This condition is generally satisfied as the plasmoids grow on the tearing mode growth time while relaxation is mostly Alfv\'enic, evidenced for example from experimental observations~\cite{janos85}. Furthermore, the relaxation time is defined with respect to a given plasmoid's local Alfv\'en time. For current sheets with many small plasmoids, the vast bulk of the plasmoid population will have the local Alfv\'en times orders of magnitude faster than the global Alfv\'en time, thereby consistent with globally super-Alfv\'enic growth rates ~\cite{pucci_velli_2014,comisso_lingam_2016,huang_comisso_bhattacharjee_2017}. The relaxation time is also much shorter than the plasmoid advection time in the current sheet or equivalently the reconnection time on order of $\sqrt{S_c}\tau_A = 100\tau_A$. Therefore it is well justified to assume that all plasmoids in the current sheet, other than those having very recently undergone coalescence, are always in a force-free state at any given time minimizing their magnetic energy while conserving magnetic helicity. For these plasmoids then, the reconnecting field flux and helicity are related according to Eq.~(\ref{helicity}), $\alpha (\zeta/B_0) \psi^3 = K$. 

As hinted above, in many instances it will be assumed that there is ample time in between mergers for a given plasmoid to reach the force-free state. Small plasmoids rarely encounter plasmoids with a lower flux, and large plasmoids experience a small change in magnetic helicity when merging, thereby not deviating significantly from the force-free state. For intermediate plasmoids, however, there may exist a regime where plasmoids merge too frequently at comparable magnetic helicities to fully relax to a force-free state in between mergers. Therefore, we develop models in two opposing limits: one model for all plasmoids to be force-free after relaxation while conserving magnetic helicity even during plasmoid mergers, as described in this subsection. The other model on non-force-free plasmoids without relaxation while conserving guide field flux, as described in the next subsection. The range of validity for the force-free assumption will require further discussion, however it will be reserved for Section \ref{sec:disc}.

Regardless of whether the resultant plasmoid is force-free, magnetic helicity is conserved if the helicity change or dissipation in the secondary current sheet during coalescence is negligible. The change in magnetic helicity during plasmoid merging is due to the change in the linkage between the guide field flux contained in the secondary current sheet and the reconnecting field flux around the current sheet. An estimate of the rate of change in magnetic helicity per change in magnetic energy, $W$, is given by \cite{ji_1999}
\begin{equation}
    \biggl| \frac{W dK}{K dW} \biggr| = 2\frac{\delta_s |B_g|}{L_s B_{rec}},
\end{equation}
where $\delta_s$ and $L_s$ are thickness and length of the secondary current sheet, respectively. Since $\delta_s/L_s$ is the steady state reconnection rate for the plasmoid merging, it should range from 0.1 to $1/\sqrt{S_c}=0.01$, depending on collisionality. Therefore, when the guide field is comparable to or weaker than the reconnecting field, no significant helicity is expected to be generated or dissipated by secondary current sheets. This permits our consideration of the plasmoid merger process as simultaneous Taylor relaxation. In a strong guide field, the dissipation of magnetic helicity would not be negligible and so the merging process would not be one of constant helicity. In fact, Taylor relaxation does not typically occur in this limit in the laboratory fusion plasmas and resultant profiles are generally not force-free.

With the assumptions above, the statistics of the new models can be described with modification of Eqs.~(\ref{eq:hb}) and (\ref{eq:hbv}). The non-velocity independent variable is chosen to be $K$, and the corresponding reconnection growth is given by Eq.~(\ref{eq:dkdt}). The loss term now includes advection and plasmoid mergers while a source is added to account for the resultant plasmoids which reached the $K$ in question by merging. If we denote the sources and sinks as $\Sigma S(K)$ and note that the influx of probability density is given by $f(K){dK}/{dt}$, then the conservation of particle number for a dynamic distribution yields
\begin{equation}\label{eq:Keq0}
\begin{gathered}
   \frac{d}{dt}\int f(K) dK + \int \frac{d}{dK}\biggl(f(K)\frac{dK}{dt}\biggr)dK = \int \Sigma S(K) dK.
\end{gathered}
\end{equation}
Taking the steady state and equating the integrands, the distribution of plasmoids with a velocity independent merger frequency is therefore given by
\begin{equation}\label{eq:Keq1}
\begin{gathered}
    3\gamma\left(\frac{\alpha\zeta}{B_0}\right)^{1/3} \frac{d}{dK}\biggl( K^{2/3} f \biggr) = \xi \delta(K) -\frac{N+1}{\tau_A}f \\ + \frac{1}{\tau_A}\int_0^{K/2}f(K')f(K-K')dK',
\end{gathered}
\end{equation}
where $N$ is the total number of plasmoids. The growth rate is present inside the derivative in order to ensure that, if only reconnection growth were present, $f$ is conserved. Here all mergers trigger a change in magnetic helicity, so the quantity $n_>$ of Eq.~(\ref{eq:hb}) is instead integrated from $0$ to $\infty$, becoming $N$. The merger term assumes the frequency of plasmoid mergers between $K-K'$ and $K'$ is $f(K')dK'/\tau_A$, adding $f(K-K')$ to $f(K)$, in a similar manner to Fermo et al \cite{fermo10} where plasmoid area is conserved. It is integrated to $K/2$ in order to avoid double counting. Combined with the merger loss term, this kinetic model is constructed such that magnetic helicity is conserved by plasmoid mergers. Each merger transforms the resultant plasmoid to a new magnetic helicity which, ignoring reconnection and advection terms, keeps the total helicity in the distribution constant. Similar modifications can be made to the velocity dependent model:
\begin{equation}\label{eq:Keq2}
\begin{gathered}
3\gamma\left(\frac{\alpha\zeta}{B_0}\right)^{1/3}\frac{\partial}{\partial K} \biggl( K^{2/3}  F \biggr) = \xi \delta(K)h(v) -\frac{H_t+1}{\tau_A}F  \\ 
+ \frac{1}{\tau_A}\int_0^{K/2} \int_{-\infty}^{\infty} \frac{|v-v'|}{V_A} F(K',v')F(K-K',v) dv'dK'
\end{gathered}
\end{equation}
with
\begin{equation}\label{eq:htdef}
    H_t(v) = \int_{0}^{\infty} \int_{-\infty}^{\infty} \frac{|v-v'|}{V_A} F(K',v') dv' dK'.
\end{equation}
Note that the notation of $F$ for velocity dependent distributions and $f$ for velocity independent distributions will be used from here on. The same process is used as before to modify the merger rate to conserve magnetic helicity. Once again, $H_t$ is simply $H$ of Eq.~(\ref{eq:hbv}) integrated from $0$, and the resultant velocity of a merger is assumed to be that of the larger plasmoid. A proof of the helicity conserving property of the merger terms, Eq. (\ref{eq:htdef}) and the last line of Eq. (\ref{eq:Keq2}), is presented in Appendix \ref{sec:app} by showing their first helicity moment is zero. It is easily extended to the velocity independent case as well. Both of these force-free models are solved numerically in Sec.III.

\subsection{A Non-Force-Free Model}

As an opposing limit of the force-free plasmoid model, which is more applicable in the strong guide field regime, we offer an alternative model where plasmoid internal field profiles are not necessarily force-free but total guide field flux is conserved by mergers. In this case, the kinetic equation in $\psi$ is unchanged from Huang et al. but the kinetic equations in $\phi$ are similar to those of the relaxing model in terms of $K$:
\begin{equation}
\begin{gathered}\label{eq:phidist}
    \gamma_g \frac{df}{d\phi} = \xi \delta(\phi) -\frac{N+1}{\tau_A}f + \frac{1}{\tau_A}\int_0^{\phi/2}f(\phi')f(\phi-\phi')d\phi'
\end{gathered}
\end{equation}
and
\begin{equation}
\begin{gathered}\label{eq:phivdist}
    \gamma_g \frac{\partial F}{\partial \phi} = \xi \delta(\phi)h(v) - \frac{H_t+1}{\tau_A}F + \\  
    \frac{1}{\tau_A}\int_0^{\phi/2}  \int_{-\infty}^{\infty} \frac{|v-v'|}{V_A} F(\phi',v')F(\phi-\phi',v) dv' d\phi'
\end{gathered}
\end{equation}
for constant and non-constant velocities, respectively.
Here $H_t$ is the same as Eq.~(\ref{eq:htdef}) with the substitution $\psi \rightarrow \phi$. Similar logic is employed for the source and loss terms, but the growth rate due to reconnection is modified. For a current sheet of thickness $\delta$, while $\psi$ is proportional to $B_{rec}\delta$ per unit length, $\phi$ is proportional to $ B_g\delta^2$. The resultant growth rate is then $\gamma_g = B_g\delta^2/\tau_{A,local} = B_g (l^2/S_c)(V_A/l) = B_gV_AL/NS_c$, where the local current sheet length $l=L/N$. This solution is also pursued numerically.


\section{Numerical Solutions to Guide Field Models}\label{sec:nums}

The method of Jacobi relaxation was used to find the steady state solution of the force-free model, and forward Euler time stepping was used to find the quasi-steady state solution of the non-force-free model\cite{press_vetterling_1999}. In both cases a logarithmically spaced grid was employed for the non-velocity variable, so the magnetic helicity or guide field flux conserving source term needed to make use of interpolation. The value of $f(K-K')$ (or $f(\phi-\phi')$) was found by using quadratic interpolation to $K-K'$ ($\phi-\phi'$), while $f(K')$ ($f(\phi')$) was simply evaluated at the grid points. All $K$ ($\phi$) integrations were performed with Simpson's rule adapted for the logarithmic grid. The $v$ integrations were performed using trapezoidal quadrature on a uniform grid. All integration source terms converge to at least second order. The convective gradient terms were calculated with Euler upwinding. To avoid roundoff error, the magnetic helicity solution was found as a function of $K^{1/3} \sim \psi$ which significantly lowered the number of decades spanned by the domain. When using Jacobi iteration the non-force-free solution suffered from unstable over-relaxation unless a very high precision was used, leading to the choice to iterate in time until a quasi-steady state was reached. In both cases the lower bound was fixed to determine $\xi$ and an upper bound was chosen equal to 0. The solutions are given assuming $B_g = B_{rec} = L = v_A = 1$ (where $B_{rec}$ is used for $v_A$), except for the strong guide field regime where $B_g = 100B_{rec}$ is used.

\subsection{Velocity Independent Models}

\begin{figure}[h]
\centering
\includegraphics[width=0.49
\textwidth]{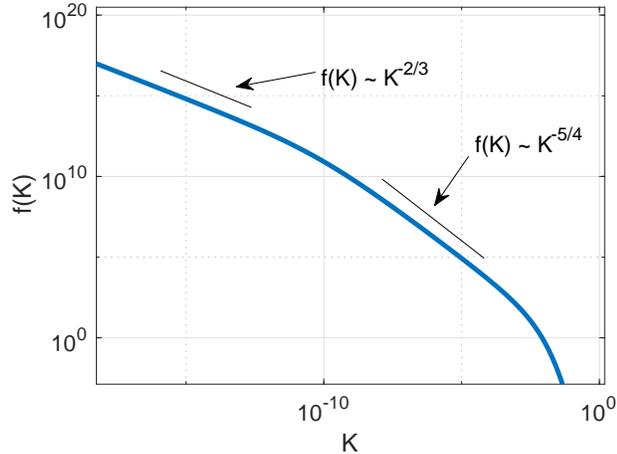}
\caption{The solution to the steady state statistical kinetic equation (Eq. \ref{eq:Keq1}) as a function of magnetic helicity, $K$, for the force-free model e with $S~10^6$. The power law slopes match the $K^{-2/3}$ and $K^{-5/4}$ lines well, with the similar exponential tail to that of Huang et al~\cite{huang_bhattacharjee_2012}, but a second power law results at low $K$ from the fact that dK/dt is not constant.}\label{fig:dist6K}
\end{figure}

\begin{figure}[h]
\centering
\includegraphics[width=0.49
\textwidth]{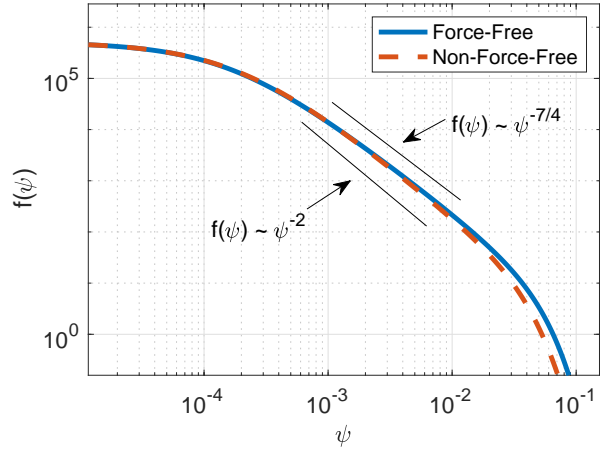}
\caption{Distributions for both the force-free (Eq. \ref{eq:Keq1} with Eq. \ref{helicity}) and non-force-free (Eq. \ref{eq:hb}) models, at $S=10^6$. There is a very slight reduction of slope in the force-free distribution with a 7/4 power law index, deviating from the $\psi^{-2}$ behavior of the non-force-free distribution.}\label{fig:distpsi}
\end{figure}

\begin{figure}[h]
\centering
\includegraphics[width=0.49
\textwidth]{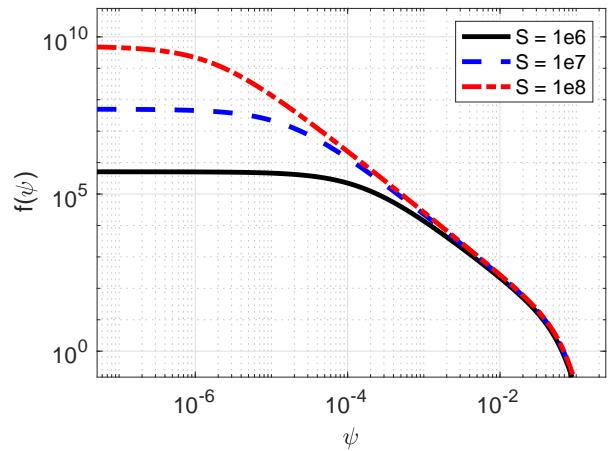}
\caption{Distribution of force-free plasmoids versus $\psi$. Results are shown for $S \sim 10^6$, $5\times 10^6$, and $10^7$.}\label{fig:distmultS}
\end{figure}

The distribution as a function of magnetic helicity for the velocity independent model was calculated for approximately $S=10^6$, which assumes $N \sim S/S_c$. It is important to stress that in practice when solving for the distributions, instead of choosing S we choose the left bound, which determines the number of plasmoids, which in turn is used to estimate S. A plot of the distribution can be found in Fig. \ref{fig:dist6K} for $N = 97.2$. The helicity dependent solution can be converted to a solution as a function of the reconnecting field flux by using $\psi = (K/\alpha(\zeta/B_0))^{1/3}$ (equivalent to assuming all plasmoids are always force-free), and enforcing $\int f_{\psi} (\psi) d\psi = \int f_K(K) dK = N$ where $dK = 3(\alpha\zeta/B_0)\psi^2 d\psi$ ($f_{\psi}$, $f_K$, and $f_{\phi}$ are simply $f$ for each variable with subscripts to emphasize their respective normalizing scale factors). The similarities between the force-free and the non-force-free models at $S=10^6$ are apparent as shown in Fig. \ref{fig:distpsi}. The exponential tail has been incrementally stretched to higher $\psi$ for the force-free model, and even more subtle is the lessening of the power law index from 2 to 7/4. These changes may in practice be hard to detect from an experimental or numerical standpoint. The flux distributions for multiple Lundquist numbers are shown in Fig. \ref{fig:distmultS} for the force-free model. As expected, at higher Lundquist numbers the power law region is extended similarly to the behavior of the non-force-free solution~\cite{huang_bhattacharjee_2012}. 

\begin{figure}[h]
\centering
\includegraphics[width=0.49
\textwidth]{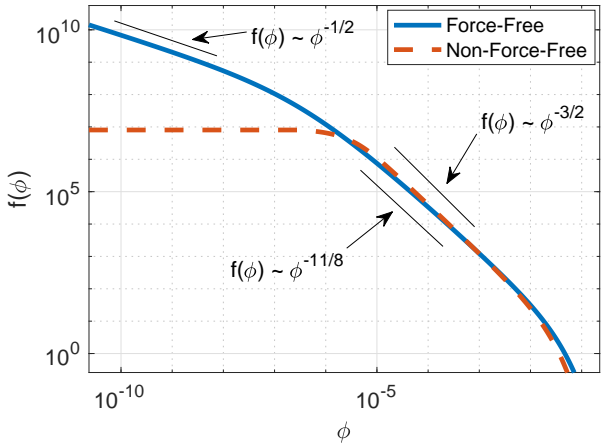}
\caption{The force-free distribution (Eq. \ref{eq:Keq1} with Eq. \ref{eqn:psiphi}) for $\sim 100$ plasmoids alongside the non-force-free distribution (Eq. \ref{eq:phidist}) for ~60 plasmoids and $B_g = 100 B_{rec}$. The non-force-free model produces a power law index of 3/2 with a constant initial region, and the relaxing model has power law indices of both 1/2 and 11/8, resulting from a non-constant $d\phi/dt$. }\label{fig:distphi}
\end{figure}

Further differences are revealed in the comparison between the force-free and non-force-free distributions as a function of $\phi$. Figure \ref{fig:distphi} shows both models for a similar number of plasmoids ($N_{ff} \sim 100$, $N_{nff} \sim 60$) with significant differences between the assumption of magnetic helicity conservation and relaxation, versus guide field flux conservation. The force-free model ties the $\phi$ distribution to the $\psi$ distribution, so the power law index of 7/4 in $\psi$ produces a power law index of 11/8 in $\phi$ because of Eq. \ref{eqn:psiphi} ($B_0 \phi = \zeta \psi^2$) and the requirement of consistent normalization ($\int f_{\psi} (\psi) d\psi = \int f_{\phi}(\phi) d\phi = N$). Additionally, the non-constant $\phi$ growth rate of the force-free model results in a second low-$\phi$ power law with index 1/2. In the non-force-free case, a steeper power law index of 3/2 is obtained, with an approximately constant initial region. Both models begin to exhibit exponential decay around roughly the same value of $\phi$.

\subsection{Velocity Dependent Models}

\begin{figure}[h]
\centering
\includegraphics[width=0.49
\textwidth]{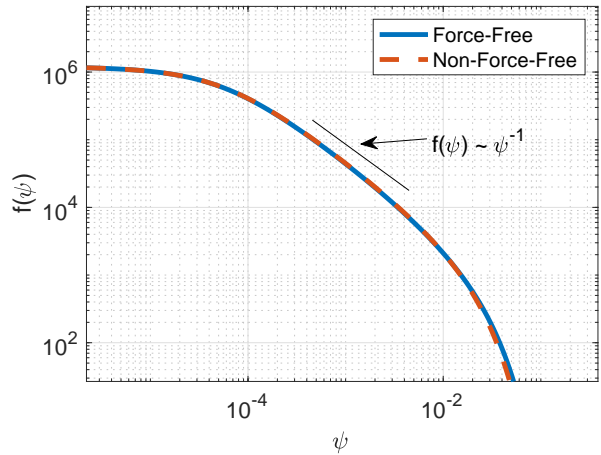}
\caption{The solutions to the steady state statistical kinetic $\psi$ equation with the velocity distribution effects included. Both the force-free (Eq. \ref{eq:Keq2} with Eq. \ref{helicity}) and non-force-free (Eq. \ref{eq:hbv}) solutions are shown with velocity dependence integrated out. The total number of plasmoids present in both solutions is $\approx 250$.}\label{fig:vdist6p}
\end{figure}

\begin{figure}[h]
\centering
\includegraphics[width=0.49
\textwidth]{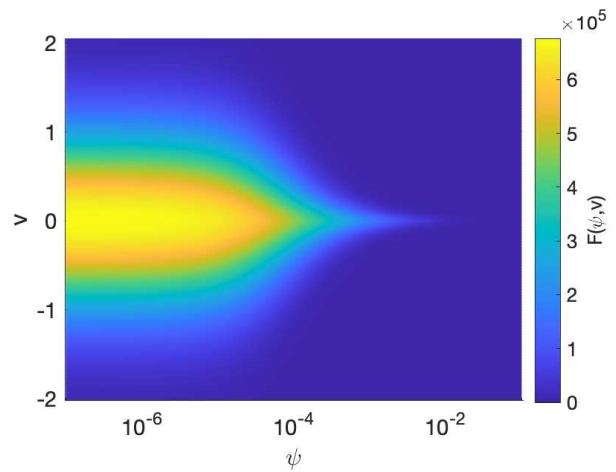}
\caption{Two-dimensional distribution showing the velocity dependence of the force-free solution. The differences between the force-free and non-force-free distributions are nuanced, with a $\sim 1.5\%$ redistribution of probability density away from the center line in the power law ($10^{-4} - 10^{-2}$) regime.}\label{fig:vdist6vp}
\end{figure}

Using the numerical methods mentioned above, the solutions to the distribution of plasmoids with varying velocities were also found. For the Gaussian velocity distribution calculated in a domain of $\psi \in [0,0.1]$ and $v \in [-3,3]$ with $f(0,0) \sim 10^6$ the solutions are shown in Figs. \ref{fig:vdist6p} and \ref{fig:vdist6vp}.

Figure \ref{fig:vdist6p} displays the solutions after translation to $\psi$ using the force-free relations, as well as after integration in $v$. It is plotted alongside the integrated non-force-free distribution for comparison. The result of the inclusion of Taylor relaxation in this model differs from that of the velocity independent model. The exponential tail is once again extended slightly and the transition to the power law region is largely unchanged, but the power law indices are both 1. Note that using $N\sim S/S_c$, the Sweet-Parker thickness $\delta$ can be found ($\delta \sim l/\sqrt{S_c} \sim L\sqrt{S_c}/S$) as $4\times 10^{-5}$. The average value of the magnetic field magnitude, $B$, for the Taylor profile is $0.575B_{rec}$, so plasmoid widths reach the order of magnitude of the current sheet thickness at $\psi = 2.32 \times 10^{-5}$ when $B$ scales linearly with $\psi$. This is relatively close to the constant pre-merger region, meaning the inability of small plasmoids to fill the current sheet may not play a significant role. Furthermore, plasmoids slightly below the current sheet width may already be sufficiently large so as to exhibit the essential properties sought for replicating the force-free behavior predicted here. Assuming the $\psi$ at which a given plasmoid first merges with another is proportional to $l = L/N = LS_c/S$, the competing processes both scale as $1/S$ and hence this balance should not change for different $S$.

\begin{figure}[h]
\centering
\includegraphics[width=0.49
\textwidth]{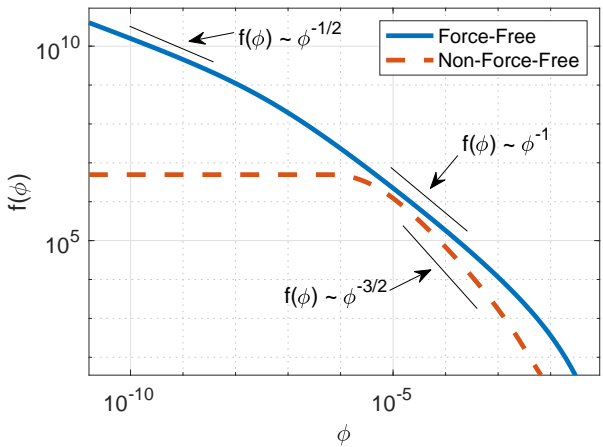}
\caption{The force-free (Eq. \ref{eq:Keq2} with Eq. \ref{eqn:psiphi}) and non-force-free (Eq. \ref{eq:phivdist}) $\phi$ distributions plotted with the velocity dependence integrated out. The non-force-free distribution has the same power law as the velocity independent case, -3/2, and the relaxing distribution exhibits a -1 power law, once again accompanied by a distinct low-$\phi$ power law with an index of 1/2. In this case the non-force-free distribution shown has $\sim 60$ plasmoids with $B_g = 100 B_{rec}$.}\label{fig:distphiv}
\end{figure}

A two-dimensional distribution of the solution in $\psi$ and $v$ is shown in Fig. \ref{fig:vdist6vp}. The power law region begins where the approximately constant distribution begins to narrow, indicating that collisions occur frequently in this region. The narrowing in the velocity spread continues until a near delta-function distribution results. The comparison of guide field distributions in the force-free and non-force-free cases are shown in Fig. \ref{fig:distphiv}. Again the magnetic helicity conserving force-free distribution presents a less steep power law than the guide field flux conserving non-force-free distribution at higher $\phi$.

\section{Discussion}\label{sec:disc}

In all models shown as functions of $\psi$, one can immediately observe similarities between the numerical solutions obtained here and those of Huang et al~\cite{huang_bhattacharjee_2012}. The initial region appears approximately constant, which can be understood as most plasmoids growing due to reconnection prior to experiencing a single merger. If $d\psi/dt = \gamma = 0.01$ and we assume the average time between mergers is $\tau_A/N$ which for 100 plasmoids is 0.01, then on average a plasmoid will not merge until they reach the size at $\psi \approx 10^{-4}$. Regarding the exponential decay, this occurs wherever the dominant loss mechanism becomes becomes advection, i.e. the remaining number of plasmoids above a certain $\psi$, $n_>$ (the integral of $f$ from $\psi$ to $\infty$), is on the order of one. This can be seen in any of the distributions by estimating $n_> \sim f(\psi) \Delta \psi$ with $\psi \sim \Delta \psi$ near the start of the exponential decay, and explains why steeper distributions transition to exponential decay at smaller $\psi$.

\begin{figure}[h]
\centering
\includegraphics[width=0.49
\textwidth]{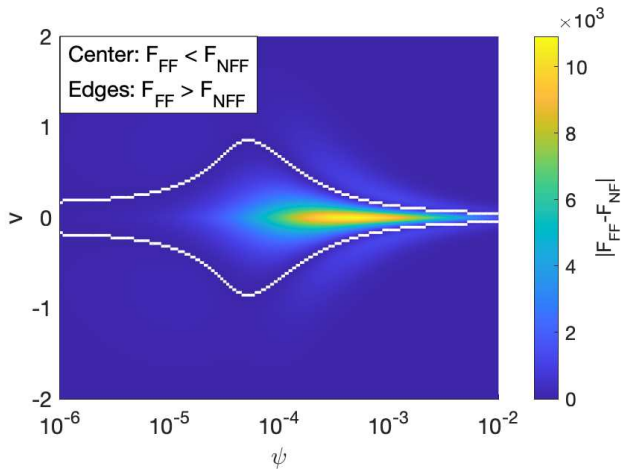}
\caption{Heat map of the difference between the force-free model and the non-force-free model. The two regions are separated based on which $F$ is larger (non-force-free "NF" and force-free "FF"). In the central low-$v$ region, $F_{NF}>F_{FF}$. In the outer regions $F_{NF}<F_{FF}$. }\label{fig:heatdiff}
\end{figure}

As can be seen in Fig. \ref{fig:distpsi}, the velocity independent force-free distribution differs only slightly from the non-force-free distribution. Of these differences in the force-free distribution, the most visible are a weak reduction in slope, and an extension of the exponential tail. The change in slope can be understood as an effect of flux conversion from guide field to reconnecting as a result of Taylor relaxation conserving magnetic helicity. Guide field fluxes add during a merger while reconnecting field fluxes do not. To satisfy the relationship $\phi = \zeta\psi^2/B_0$ [Eq.(\ref{eqn:psiphi})] for the force-free field after relaxation while conserving magnetic helicity, the excess guide field flux must be converted to reconnecting field flux. In practice, using these relationships one finds that at most 20.6\% of the guide field flux will be converted after a merger between equivalent flux plasmoids. Given that smaller plasmoids are much more populous, mergers often result in little flux conversion, and hence the slope change in $f$ is subtle. This can be understood more quantitatively by a comparison between the merger terms in the force-free and non-force-free statistical equations. These terms side by side can be rewritten for $\psi$ as (dropping the $\tau_A$):
\begin{equation}
\begin{gathered}
    \int\displaylimits_0^{\psi/2^{1/3}} \biggl(1-\frac{\psi'^3}{\psi^3} \biggr)^{-2/3} f(\psi')f(\sqrt[3]{\psi^3-\psi'^3})d\psi' - Nf(\psi) \\
    \textrm{and} \qquad \int\displaylimits_0^{\psi}f(\psi)f(\psi')d\psi' - Nf(\psi)
\end{gathered}
\end{equation}
where a scaling factor is present in the force-free expression as a result of the transition from $K$ to $\psi$. Due to the monotonic decreasing nature of $f$, $f((\psi^3-\psi'^3)^{1/3}) \geq f(\psi)$ over the bounds of integration, increasing the integrand, and the overall value of the source in Eq.~(\ref{eq:Keq1}). This is further aided by the scaling factor, which within the bounds of the integral reaches a maximum of $2^{2/3}$. The integration interval is slightly shortened by the upper bound of $\psi/2^{1/3}$, however due to the steepness of $f$ and the scaling factor, the latter effect is outweighed by the former. In the case of the velocity dependent model the slope is far less steep and the collision rate is weighted by differential velocity (see Fig.~\ref{fig:vdist6p}). This more mild slope lowers the enhancement of $F((v,\psi^3-\psi'^3)^{1/3})$ in the integral, allowing it to be muted by an effect introduced with the inclusion of a velocity distribution. A comparison of the $\psi-v$ dependence of the non-force-free distribution with the force-free distribution in Fig.~\ref{fig:heatdiff} shows that in the power law regime, the force-free distribution experienced an outward shift of plasmoids from the central low $v$ region to higher $v$ regions. Overall, the average speed of plasmoids increased by 1.5\%. While this change is small, this causes plasmoid mergers to occur more frequently in the power law region, increasing the steepness of $F$ slightly. This may be an effect of flux conversion leading to more rapidly moving low flux plasmoids "leapfrogging" into the merger-dominated regime and shortly merging with other plasmoids. While this bolstered collisionality may increase the steepness, any merger resulting in flux conversion still boosts the $\psi$ of one of the plasmoids. That is why in both the velocity independent and dependent force-free models, the exponential tail of the distribution is extended, albeit to a lesser extent in the velocity dependent model.

From Fig.~\ref{fig:distphi}, the differences in the velocity independent guide field distributions are much more pronounced than those of $\psi$. While one may expect flux conversion of guide field flux to reconnecting flux would cause the force-free model to have a steeper slope than the non-force-free model (at least without velocity), the results are quite the opposite. This is due primarily to the reconnection growth rates of $\phi$ in the two models. In the non-force-free case the growth rate is the constant $\gamma_g$ for all plasmoids, however, the relaxing model does not grow $\phi$ at a constant rate. The nature of the guide field flux in this model is assumed to be passive, and a plasmoid's $\psi$ grows in a manner which maintains a force-free state at the expense of the growth of guide field flux. Therefore the plasmoid follows $\phi = \zeta\psi^2/B_0$ [Eq.(\ref{eqn:psiphi})] so that $d\phi/dt = 2\gamma\sqrt{B_0 \phi/\zeta}$, growing faster the larger it gets. The force-free growth rate, although smaller than $\gamma_g$ initially, surpasses its non-force-free counterpart near $\phi \approx 10^{-6}$. This non-constant growth rate leads to another distinguishing feature of the force-free model, a 1/2 power law index at low $\phi$. This two versus one power law difference between models is much more significant than that which occurs later in $\phi$, and could significantly ease the difficulty in determining which model is most appropriate for a given data set. The same effects are seen in the velocity dependent distribution comparison, although in this case the non-force-free power law is the same as the velocity independent version, even though the force-free power law has changed. This relates to the increase in steepness observed in the velocity dependent force-free $\psi$ distribution. Since $\phi \sim K^{2/3}$ in the force-free model, mergers add $\phi$ in a sort of two-thirds quadrature. However, directly adding $\phi$ in the non-force-free model allows for a larger "jump" when merging, leading to a more significant difference in slopes.

\begin{figure}[h]
\centering
\includegraphics[width=0.49
\textwidth]{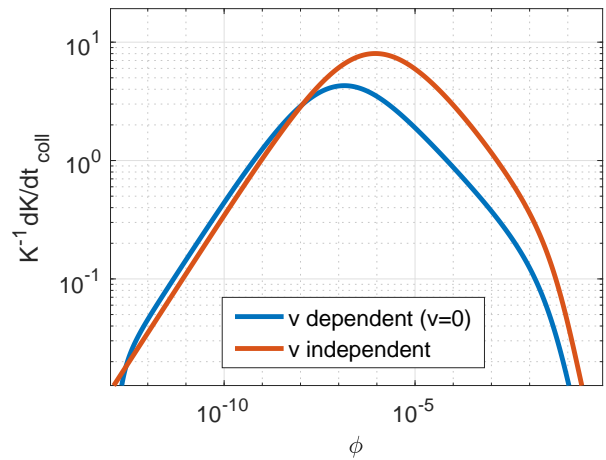}
\caption{Plot which shows rate of change of $K$ due to mergers, normalized to the given $K$ as a function of $\phi$. It demonstrates where the average change in helicity due to a merger is small enough that a plasmoid may be able to partially or fully relax in between collisions. }\label{fig:krate}
\end{figure}

In Section \ref{sec:dist}, the force-free model was chosen to be the limit where all plasmoids are in a force-free state at all times, even when merging. However, it is possible that a plasmoid which undergoes two rapid consecutive mergers may not reach a force-free state in between the mergers, even though magnetic helicity conservation is unaffected. Figure \ref{fig:krate} shows the rate of change of a plasmoid's magnetic helicity, normalized to its helicity right before merging as a function of $\phi$, solely due to mergers for the force-free models. This reveals where in the distribution a given plasmoid experiences the most frequent and significant mergers. The velocity dependent model is shown evaluated at $v=0$. As an exercise, if one considers relaxation to be possible when the helicity changes by no more than the given $K$ in an Alfv\'en time, the $v$-independent model would exhibit force-free plasmoid configurations before $\phi \approx 10^{-9}$ and after $\phi \approx 10^{-3}$, while slower plasmoids in the $v$-dependent model would be force-free after $\phi \approx 10^{-4}$ as well as before $\phi \approx 10^{-9}$. The drop off at high $\phi$ indicates that the largest plasmoids in the exponential tail should be frequently found in a relaxed state. However, this is simply meant to illustrate where in the distribution plasmoids undergo the most change. If relaxation can take place sufficiently rapidly, then one will observe characteristics of the force-free model in the region of rapid helicity change. Lastly, we note that we assume that mergers of 3 or more plasmoids are rare enough so that their effects on the distribution are negligible.


\section{Conclusions}\label{sec:conc}

These models do not necessarily represent the exact distributions of plasmoids in a current sheet which undergoes dynamic reconnection. Due to the disruption of relaxation by frequent mergers at some scales, it may be more appropriate to treat the non-force-free and force-free cases as asymptotic behaviors. From Fig.~\ref{fig:krate}, we expect to observe force-free plasmoids at the smaller scales, as well as the largest. The suspicion that large $\psi$ plasmoids may be able to relax is hinted at by the observation of Taylor-like relaxed flux ropes present in the solar wind as well as the earth's magnetotail~\cite{sun_2019,telloni_2016}. While in the Hall-MHD regime, these results may still indicate that larger plasmoids are capable of achieving a force-free state, suggestive of some processes dominated by magnetic helicity conservation.

The results here indicate that guide field flux distributions may be noticeably distinct, especially due to the introduction of a second low-$\phi$ 1/2 power law index in the force free model. Experimentally, however, this can be a challenging quantity to measure. A more directly measurable characteristic would be the distribution of plasmoid sizes, or the quantity $a$ in the force-free case. From the surface integration of the reconnecting field we can easily demonstrate that $\psi \sim a$ in the force-free case. Unfortunately, the distinction of this from the non-force-free model may also prove difficult. Using the conclusions of Uzdensky et al \cite{uls_2010}, in the incompressible limit of the anti-parallel plasmoid instability the width of a plasmoid perpendicular to the current sheet would follow $w_p \sim \psi/B_{rec}$. This incompressible limit may also be thought of as the case of strong guide field where the added magnetic pressure augments the pressure of the magnetized plasma within. Hence one may only be able to distinguish the physical size distributions as well as the $\psi$ distributions can be distinguished between the force-free and non-force-free models. Instead, a more accessible indication that helicity conservation is, to some extent, playing a role may simply be the presence of an enhanced core field in line with that of the Taylor profiles internal to the plasmoids which form the backbone of the force-free model~\cite{sun_2019,taylor_1986,slavin95,Slavin03jgr}. It should be noted that if the $\psi$ distributions (or more importantly $a$ or $w_p$ distributions) are similar between the force-free and non-force-free models and the $\phi$ distributions vary largely, then the internal plasmoid guide field pressures will also be significantly different. Here, guide field pressure would reach noticeably greater levels in the force-free model over the non-force-free model.

The models presented here may also prove relevant to the plasmoid distribution when the external guide field is negligible. Instabilities which arise in three dimensions such as kink modes add an out-of-plane magnetic field to an otherwise anti-parallel configuration. The process of flux rope merging in 3D is a particularly important example of an event where kinking occurs in an otherwise straight cylindrical island in 2D. The condition of thin flux ropes being necessary should be mentioned to ensure that the characteristic length scale of a kink does not largely affect the assumption of approximately straight cylindrical Taylor states (as in high aspect ratio reverse field pinch devices). This negligible guide field regime is not forbidden in any way by our equations. Rather, it is the case where the quantity $\lambda a$ is the first zero of $J_0(x)$. This phenomenon of Taylor relaxation where the external guide field is zero has been observed experimentally\cite{craig17}.

The addition of relaxation has also allowed for an increase in the spread of plasmoid velocities. Although it is a small change of $1.5\%$, a Fermi-like process involving multiple reflections from plasmoids can add up to cause an increase in the highest achievable energy of charged particles undergoing this acceleration. More significantly, the acceleration which occurs during island contraction relies on how quickly the island is compressed. The enhancement of guide field pressures in the relaxing model and the increasing rate with which guide field flux is accreted onto a plasmoid can result in a greater effective mirror velocity in a first order process~\cite{2020spitkovsky}.

The force-free model's alterations to particle acceleration processes and the speed of plasmoids in the outflow are unlikely to affect the global rate of energy conversion in any considerable capacity. While energization may be enhanced for some high energy particles already capable of undergoing Fermi acceleration, they represent a minute fraction of the overall energy of the plasma and therefore are not likely to contribute to the global energy conversion rate in a significant manner. It is possible, however, that dissipation of magnetic energy during the Taylor relaxation that accompanies plasmoid growth and mergers could result in a modified conversion rate. The assumption of perpetual relaxation as a plasmoid grows due to reconnection leaves the reconnecting flux growth rates the same, but the guide field flux growth rates differ significantly between models. In addition it can be shown that a force-free plasmoid's total magnetic energy goes linearly with its guide field flux. For large plasmoids, the guide field flux growth rate is enhanced in the force-free model, but for small plasmoids it is significantly hindered. Because of this it is not clear whether the overall rate of energy conversion would increase or decrease between models. Therefore we leave this calculation to future investigations.

In order to make a more robust model of the plasmoid distribution in a guide field, several more effects may be included in the future. Specifically, the assumption of merged plasmoids adopting the velocity of the larger allows for the narrowing of the velocity dependent distribution to a near delta function in $v$. If instead flux weighted averaging was used, a greater spread in velocity at higher $\psi$ would result. This broadening is also suggested by the spread in the velocity distribution found by Lingam and Comisso~\cite{lingam_comisso_2018}. In parallel, a more thorough investigation of the effects of the choice of $H(v)$ on the distribution of plasmoid velocities may be of interest. While this work has focused on the distribution in fluxes and helicity, more realistic merging rules may be accompanied by varied effects on the distribution of plasmoid velocities. It may also prove important to address the possibility of incomplete or particularly slow plasmoid mergers. Coalescence rates have been observed to stall in simulations due to a sloshing effect at high Lundquist numbers~\cite{knoll2006}. This delayed or inhibited coalescence could give rise to more efficient production of high flux plasmoids in our statistical models. Additionally, there have been growth rates proposed for the plasmoid instability which would modify the Sweet-Parker reconnection rate $\gamma$ used here. While they may not affect the power law slope, an increase or decrease in $\gamma$ would affect where the transition from the constant region to the power law region occurs. Numerical or experimental investigations of these distributions may seek to determine which model is most appropriate for the problem of plasmoid unstable reconnection in a guide field, or whether they both remain contained within their expected regimes. Observation of power law behavior outside the limits of the force-free and non-force-free models may suggest that conservation of an alternative quantity holds over guide field flux or magnetic helicity, while something in between may, if appropriate, suggest an intermediate regime which possesses characteristics of both distributions.





\appendix

\section{Magnetic Helicity Conservation of the Collision Terms in the Plasmoid Kinetic Equation }\label{sec:app}

A required characteristic of our plasmoid kinetic equation is that the terms involving plasmoid mergers must conserve magnetic helicity in the distribution. In other words, the first moment of the following equation, from Eq.~(\ref{eq:Keq2}), must be zero:
\begin{multline}\label{eq:Keq3}
 \biggl(\frac{dF}{dt}\biggr)_{merge} = -\frac{F}{\tau_A} \int\limits_{0}^{\infty} \int\limits_{-\infty}^{\infty} \frac{|v-v'|}{V_A} F(K',v') dv' dK' \\
+ \frac{1}{\tau_A}\int\limits_0^{K/2} \int\limits_{-\infty}^{\infty} \frac{|v-v'|}{V_A} F(K',v')F(K-K',v) dv'dK'.
\end{multline}

In this section we will prove that this is true. We begin by integrating the equation over the entire velocity space $v$. For readability, the we will use the shorthand $\int\limits_{v} = \int_{-\infty}^{\infty} dv$ and $\int\limits_{K} = \int_{0}^{\infty} dK$, as well as $\Delta \hat{v} = |v-v'|/V_A$:


\begin{equation}\label{eq:Keq4}
\begin{gathered}
 \int\limits_{v} \biggl(\frac{dF}{dt}\biggr)_{merge} = -\frac{1}{\tau_A} \iiint\limits_{v,v',K'} \Delta \hat{v} F(K,v) F(K',v') \\
+\frac{1}{\tau_A} \iint\limits_{v,v'} \Delta \hat{v} \int\limits_0^{K/2} F(K',v')F(K-K',v) dK'.
\end{gathered}
\end{equation}

Some reshuffling of the final term will prove useful. First, a substitution of $x = K-K'$ yields


\begin{equation}
\begin{gathered}
    \iint\limits_{v,v'} \Delta \hat{v} \int\limits_0^{K/2} F(K',v')F(K-K',v) dK' \\ =
    \iint\limits_{v,v'} \Delta \hat{v} \int\limits_{K/2}^{K} F(K-x,v')F(x,v) dx.
\end{gathered}
\end{equation}

Given that $v$, $v'$, and $x$ are integrated out, we can rename them as $v'$, $v$, and $K'$, respectively, without affecting the results

\begin{equation}
\begin{gathered}
    \iint\limits_{v,v'} \Delta \hat{v} \int\limits_0^{K/2} F(K',v')F(K-K',v) dK' = \\
    \iint\limits_{v',v} \Delta \hat{v} \int\limits_{K/2}^{K} F(K-K',v)F(K,v') dK'.
\end{gathered}
\end{equation}

Therefore, the identical integrand on both side of the above equation permits the following relationship:

\begin{equation}
\begin{gathered}
    \iint\limits_{v,v'} \Delta \hat{v} \int\limits_0^{K/2} F(K',v')F(K-K',v) dK' = \\
    \frac{1}{2} \iint\limits_{v,v'} \Delta \hat{v} \int\limits_0^{K} F(K',v')F(K-K',v) dK'.
\end{gathered}
\end{equation}

A property of our distributions is that $F(K,v) = 0$ for $K<0$ and equivalently $F(K-K',v) = 0$ for $K'>K$. Therefore we may extend the $K$ integration out to infinity producing a convolution, and replace the result into our integrated merger terms:

\begin{equation}\label{eq:Keq5}
\begin{gathered}
 \int\limits_{v} \biggl(\frac{dF}{dt}\biggr)_{merge} = -\frac{1}{\tau_A} \iiint\limits_{v,v',K'}  \Delta \hat{v} F(K,v) F(K',v') \\
+\frac{1}{2\tau_A} \iint\limits_{v,v'} \Delta \hat{v} F(v')*F(v),
\end{gathered}
\end{equation}

where we use the standard notation for a convolution $F(v')*F(v)$ with explicit $v$ dependence to differentiate between the primed and unprimed variables. The next step is to Laplace transform this equation from $K$ to $s$, and take a derivative with respect to $s$:

\begin{equation}\label{eq:Keq6}
\begin{gathered}
 -\iint\limits_{v,K} K e^{-sK} \biggl(\frac{dF}{dt}\biggr)_{merge} = \\
 - \frac{1}{\tau_A}  \iint\limits_{v,v'} \Delta \hat{v} \mathcal{L}'[F](s,v) \mathcal{L}[F](0,v') \\
+ \frac{1}{2\tau_A} \iint\limits_{v,v'} \Delta \hat{v}  \biggl[\mathcal{L}[F](s,v') \mathcal{L}'[F](s,v) \\
+  \mathcal{L}'[F](s,v') \mathcal{L}[F](s,v) \biggr].
\end{gathered}
\end{equation}

Taking the negative of Eq.~(\ref{eq:Keq6}) and evaluating it at $s=0$,

\begin{equation}
  -\iint\limits_{v,K} \mathcal{L'}\biggl[\biggl(\frac{dF}{dt}\biggr)_{merge}\biggr](0,v) = 
    \iint\limits_{v,K} K \biggl(\frac{dF}{dt}\biggr)_{merge} = 0.
\end{equation}
This proof can easily be generalized to the velocity independent case, and is also of course applicable to the guide field conserving regime with $K$ replaced with $\phi$. 

\medskip
The data that support the findings of this study are available from the corresponding author upon reasonable request.

\begin{acknowledgments}
This work is supported by the DOE through contract No. DE-AC0209CH11466. The authors would like to thank Yi-Min Huang for his help with numerical methods, careful verification of results, and correction of mistakes, as well as William Fox for his helpful feedback and Amitava Bhattacharjee for his valuable conceptual discussions.
\end{acknowledgments}

\bibliography{plasmoids.bib,rec.bib}

\end{document}